\documentclass[11pt, conference]{IEEEtran}
\usepackage[margin=.9in,nohead,nofoot]{geometry}
\usepackage{subfigure}
\usepackage{epsfig,graphicx,psfrag}
\usepackage{amsfonts,amsmath,amssymb}
\usepackage{multirow}

\newtheorem{theorem}{Theorem}[section]

\newtheorem{corollary}[theorem]{Corollary}

\begin{document}

\title{Coordination using Implicit Communication}

\author{
\small
\authorblockN{Paul Cuff}
\authorblockA{Princeton University\\
cuff@princeton.edu }
\and
\authorblockN{Lei Zhao}
\authorblockA{Stanford University\\
leiz@stanford.edu}
}

\maketitle

\begin{abstract}
We explore a basic noise-free signaling scenario where coordination and communication are naturally merged.  A random signal $X_1,...,X_n$ is processed to produce a control signal or action sequence $A_1,...,A_n$, which is observed and further processed (without access to $X_1,...,X_n$) to produce a third sequence $B_1,...,B_n$. The object of interest is the set of empirical joint distributions $p(x,a,b)$ that can be achieved in this setting.  We show that $H(A) \geq  I(X;A,B)$ is the necessary and sufficient condition for achieving $p(x,a,b)$ when no causality constraints are enforced on the encoders.  We also give results for various causality constraints.

This setting sheds light on the embedding of digital information in analog signals, a concept that is exploited in digital watermarking, steganography, cooperative communication, and strategic play in team games such as bridge.
\end{abstract}

\section{Introduction}

We are interested in examining a simple batch of communication questions that obscure the line between ``analog'' control and ``digital'' communication signaling.  How well can a signal be used to both carry information (digital) and play an explicit role in a system (analog)?  Suppose a communication signal is required to have certain statistical properties and correlations with other signals of interest, such as in a multiuser communication setting, or consider a control signal that is used to carry additional embedded information.  This sort of dual purpose signaling manifests itself naturally in the simple communication setting shown in Figure \ref{figure setup}.

\subsection{An ``Online Communication'' Problem}

Let us begin the discussion with an example from the literature.  In 2003, Gossner et. al. \cite{Gossner03} solved an interesting problem involving sequential play of a cooperative penny matching game.  The game setting allows for communication between the players only through actions in the game, which they refer to as ``online communication.''  The game involves a random binary sequence (the ``source'') and two players, Alice and Bob.  Alice knows the source sequence, but Bob doesn't.  Alice and Bob repeatedly attempt to guess\footnote{Alice knows the source sequence, so her ``guesses'' are always correct if she chooses.  The optimal strategy will have Alice inserting wrong guesses for the sake of communication.} the source sequence, one bit at a time.  They obtain one point whenever both of them guess correctly.  After each guess, they each see the guess of the other person and the source bit.  As you might expect, they are allowed to strategize before the source sequence is revealed to Alice, but after the game begins they cannot communicate explicitly -- only implicitly through the game itself.  What is the best average score that can be achieved?

Gossner et. al. show that the optimal average score of this game is .82, which is significantly better than the average score that can be achieved through trivial (albeit clever) strategies.  \emph{(Warning: Spoiler!  Pause here if you wish to solve this problem on your own.)}  You can achieve this score using techniques from communication theory (error-correction codes) and information theory.  The main ideas are block-Markov coding, rate-distortion theory for Hamming distortion, and input-constrained channel capacity (binary channel with no noise).  The analysis by Gossner et. al. was combinatoric instead of information theoretic.  They also present a matching upper bound which is very specific to the particular game being played.

A nice surprise related to this game emerges from the results of our work.  Suppose that the game was made more difficult.  After each guess, Bob sees the guess that Alice made but does not see the source bit (nor does he know the score of the game until after the game is finished).  It turns out that the optimal average score of the game is the same!  This may be surprising because the strategy prescribed by Gossner et. al. to achieve optimality requires that Bob consider the past source bits when making his next guess.  The strategy must be significantly modified in order to achieve optimality when Bob does not see the past source bits.  This observation is not limited to the specific repeated game being played. We provide an information theoretical solution to general games of this form in Section~\ref{section causal}.

\section{Uses and Illustrations}

We encounter a variety of situations in signal processing and communication where a signal plays multiple roles.  Perhaps the most relevant to this work are those involving network communication.  In a multiuser joint source-channel coding setting, the encoders must structure communication signals to convey information about the sources while also taking advantage of statistical dependencies of the sources to correlate and align the communication signals.

A specific situation where a communication signal is used directly and indirectly is the ``cribbing'' transmitters encountered in the work of Van der Meulen \cite{vanderMeulen77} and Willems \cite{Willems-vanderMeulen85} \cite{Willems82}, and more recently by Permuter and Asnani \cite{Permuter10}.  Here a multiple access channel is considered, but the channel input from one transmitter is overheard by the other transmitter, allowing them to learn about each other's message and cooperate.  Here it is discovered that the channel input should not only carry information intended directly for the other transmitter, but it should also be a suitable transmission signal.

In other examples, there are explicit goals to embed information in signals, such as digital watermarking and steganography.  Here, a media signal, such as video or audio, is augmented to carry information in the form of an ID tag or data, which is usually intended to be imperceivable to human perception.  Research exploring the capacity to embed information under signal distortion constraints can be found in \cite{Wu-Hwang07}, \cite{Chen-Wornell00}, \cite{Cohen-Lapidoth02}, and \cite{Moulin-O'Sullivan03}.

Let us now suggest some illustrations of the scenario we are concerned with in a concrete, though playful, manner.

\subsection{Game of Bridge}

In the game of bridge, players bid for contracts which allow them to call trump, pass cards, and hopefully earn enough points to validate the contract.  The bid consists of a number and a suit, indicating how many tricks will be won (beyond the defacto six) and a suit for trump.  However, a player who makes a first bid of `1 Clubs' may not be bidding for the sake of winning the contract.  Instead, the bid might be a message to his partner that there is no dominant suit in his hand.  Communication strategies for bridge are limited by the effect they have on the play of the game.

\subsection{Collusion}

High speed stock trading systems make money by their precise timing of buying and selling.  Suppose two trading systems wish to collude in order to shift market prices, and they wish to do so in a way that is not discoverable over standard communication channels.  How much can they communicate through the timing of their buys and sells without adversely affecting their profits?

\subsection{Multi-part Printing}

Two printers are used to print a color document.  The first prints all colors, and the second prints black only.  However, the electronic document for printing is sent only to the first printer.  The second printer scans the color document and adds black where needed.  The color printer, which mixes three inks to create black, can save ink by leaving black for the second printer to take care of, but information about the location of the black must be written into the image somehow.  How much ink can be saved?

\section{Cascade of Controllers}
\label{section setup}

\subsection{Problem Statement}

An i.i.d. random process $\{X_i\}$ is distributed according to $p_X$, which is to say that any finite block of symbols is distributed according to
\begin{eqnarray*}
X^n & \sim & p_{X^n}(x_1,...,x_n) \; = \; \prod_{i=1}^n p_X(x_i).
\end{eqnarray*}

\begin{figure}
\psfrag{a}[][][0.8]{$X^n$}
\psfrag{b}[][][0.8]{$A^n$}
\psfrag{c}[][][0.8]{$B^n$}
\psfrag{d}[][][0.8]{Controller 1}
\psfrag{e}[][][0.8]{Controller 2}
\psfrag{f}[][][0.8]{Source}
\psfrag{g}[][][0.8]{Actuator 1}
\psfrag{h}[][][0.8]{Actuator 2}
\centering
\includegraphics[width=.5\textwidth]{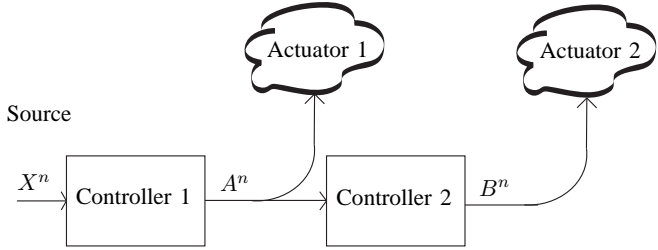}
\caption{{\em Cascade of controllers.}  The source of information, $X^n$, is an i.i.d. sequence with a known distribution.  Controller 1 produces a control sequence $A^n$ which has information embedded into it for Controller 2.  Without access to the source, Controller 2 processes $A^n$ to produce a control sequence $B^n$.}
\label{figure setup}
\end{figure}

The cascade of controllers shown in Figure \ref{figure setup} produces two additional sequences $\{A_i\}$ and $\{B_i\}$.  The $A_i$'s are a function of the $X_i$'s and the $B_i$'s are a function of the $A_i$'s, possibly with causality constraints.  The system runs for a finite but arbitrarily large number of iterations, $n$, and we use the superscript notation $X^n$ to represent the sequence $X_1,...,X_n$.  The goal is to coordinate the sequence of triples $(X,A,B)_i$ with a desired empirical distribution.

We characterize the coordination that is achievable among the three control signals in terms of the empirical coordination of \cite{Cuff-Permuter-Cover10}.  Under this framework, a coordination scheme is summarized by the joint distribution that it achieves, in the sense that the frequencies of triples $(X,A,B)_i$ correspond closely with the specified joint distribution with high probability.  Unlike the problems considered in \cite{Cuff-Permuter-Cover10}, the cascade of controllers setting has no explicit rate-limited communication channels.

To state the criterion for empirical coordination formally, a conditional distribution $p(a,b|x)$ can be achieved if for all $\epsilon>0$ there exists an integer $n$ and encoding functions $f$ and $g$ (satisfying the necessarily causality constraints) such that
{\small
\begin{eqnarray*}
\mathbf{P}\left( \left\| P_{ X^n,A^n,B^n }(x,a,b) - p_0(x) p(a,b|x) \right\|_{TV} > \epsilon \right) & < & \epsilon,
\end{eqnarray*}}
where  $A^n \in {\cal A}^n$, $B^n \in {\cal B}^n$, the induced empirical distribution
$P_{X^n,A^n,B^n}(x,a,b)= \frac{1}{n} \sum_{i=1}^n \mathbf{1}_{\{(X_i,A_i,B_i) = (x,a,b)\}}$,
and $\| \cdot \|_{TV}$ is the total variation distance between two distributions.

The coordination set of all achievable distributions for empirical coordination is designated as
\begin{eqnarray*}
{\cal P} & \triangleq & \left\{ \mbox{achievable } p(a,b|x) \right\}.
\end{eqnarray*}
The main results of this paper are the characterizations of the coordination sets in Theorem~\ref{theorem noncausal}, Theorem~\ref{theorem causal}, and Figure~\ref{fig.d1d2}.

\subsection{Sequences - An Alternative Statement}

The coordination scenario of this paper is described as controllers acting on signals, providing a natural operational meaning.  However, the results of the analysis in this work are simply statistical and probabilistic statements about sequences. Consider the set of all groups of random variables $X^n$, $A^n$, and $B^n$ having the following two properties.  First, $X^n - A^n - B^n$ forms a Markov chain.  Second, $X^n$ is an i.i.d. sequence according to $p_0(x)$.  This is exactly the set of random variables that can be produced by a cascade of non-causal randomized controllers.  Theorem \ref{theorem noncausal} then relates to the first-order statistics of the sequences in this set.

\subsection{Maximize Average Score}
\label{subsection average score}

We can take a different approach to analyzing coordination by specifying a reward function for the three combined signals.  Let the function $\Pi(x,a,b)$ be a reward obtained for each occurrence of the triple $(x,a,b)$ in the sequence of combined signals $(X,A,B)_1, (X,A,B)_2,...$.  We can then ask for the greatest possible average reward under the constraints imposed by the cascade of controllers of Figure \ref{figure setup}, taking the supremum over all choices of block length $n$ and controllers.

It turns out that this analysis is fundamentally the same as characterizing the coordination set ${\cal P}$.  The optimal average reward corresponding to the function $\Pi$ can be found by maximizing $\mathbf{E} \; \Pi (X,A,B)$ over the coordination set of conditional distributions.  Likewise, the coordination set, being a convex set, is fully characterized by the optimal average reward for all reward functions $\Pi$.  This connection is due to the close relationship between the average function value of a sequence and the empirical distribution.  For a detailed proof of the relationship, see the discussion in Section VI of \cite{Cuff-Permuter-Cover10} and the proof in Section VII.

\section{Non-causal Controllers}
\label{section noncausal}

Controller 1 and Controller 2 produce signals according to unconstrained non-causal encoding functions:
\begin{eqnarray*}
A^n & = & f(X^n), \\
B^n & = & g(A^n).
\end{eqnarray*}

\begin{theorem}
\label{theorem noncausal}
The coordination set ${\cal P}$ for the cascade of controllers in Figure \ref{figure setup} is the set of conditional distributions $p(a,b|x)$ such that the joint distribution with the source, given by $p_0(x)p(a,b|x)$, satisfies
\begin{eqnarray*}
H(A) & \geq & I(X;A,B).
\end{eqnarray*}
\end{theorem}

\subsection{Achievability}

To efficiently achieve coordination with a cascade of controllers, we populated a codebook of $(a^n,b^n)$ pairs.  Controller 1 identifies a pair $(\tilde{a}^n, \tilde{b}^n)$ in the codebook which yields the desired correlation with $X^n$.  However, Controller 1 only produces $A^n = \tilde{a}^n$, which is the first half of the codeword.  If the codebook is small enough, Controller 2 will be able to identify which codeword Controller 1 selected based only on observing $A^n$.

Consider a source distribution $p_0(x)$ and a desired conditional distribution $p(a,b|x)$ that satisfies $H(A) > I(X;A,B)$.  Select a constant $r$ such that $H(A) > r > I(X;A,B)$.  Let ${\cal C} = \{(a^n(k),b^n(k))\}_{k=1}^{2^{nr}}$ be a randomly generated codebook, where each $(a^n(k),b^n(k))$ is independently drawn from the marginal distribution induced by $p_0(x) p(a,b|x)$.

Controller 1 finds an integer $k$ such that $(X^n,a^n(k),b^n(k))$ is jointly typical (in the sense that the empirical joint distribution is close to the desired distribution in total variation).  This will be successful with high probability if $n$ is large enough, as a consequence of rate-distortion theory, since $r > I(X;A,B)$.  Controller 2 searches the codebook ${\cal C}$ for the first $j$ such that $a^n(j) = A^n$ and produces the control sequence $B^n = b^n(j)$.  If Controller 1 was successful, then $A^n$ is a typical sequence, and with high probability there is no other codeword in the randomly generated codebook equal to $A^n$ since $r < H(A)$.

\subsection{Converse}
This problem does not involve rates of communication.  The converse rests on the following observation.
{\allowdisplaybreaks
\begin{eqnarray*}
H(A^n) & \geq & I(X^n;A^n) \\
& \stackrel{(a)}{=} & I(X^n;A^n,B^n),\\
& = & \sum_{q=1}^n I(X_q;A^n,B^n|X^{q-1}) \\
& = & n I(X_Q;A^n,B^n|X^{Q-1},Q) \\
& = & n I(X_Q;A^n,B^n,X^{Q-1},Q) \\
& \geq & n \; I(X_Q;A_Q,B_Q).
\end{eqnarray*}}
where (a) comes from the fact that $X^n-A^n-B^n$ form a Markov chain. $Q$ is a time sharing random variable uniformly distributed on $\{1,...,n\}$ and independent of $\{X^n,A^n,B^n\}$. Similarly,
\begin{eqnarray*}
H(A^n) & = & \sum_{q=1}^n H(A_q|A^{q-1}) \\
& = & n \; H(A_Q|A^{Q-1},Q) \\
& \leq & n \; H(A_Q).
\end{eqnarray*}

\section{One Causal Controller}
\label{section causal}

Let us revisit the game Gossner et. al. solved in \cite{Gossner03}. In their setting, Controller 1 observes the whole $X^n$ sequence and then generates an action sequence $A^n$\footnote{Technically, Controller 1 also observes $B_1,...,B_{i-1}$ when producing action $A_i$, but this can be safely ignored.}; Controller 2 has a sequence of causally constrained action functions $g_i(\cdot)$ for $i=1,...,n$. Therefore, the controllers act according to the following encoding functions:
\begin{eqnarray*}
A^n & = & f(X^n), \\
B_i & = & g_i(A^{i-1}) \mbox{ for } i = 1,...,n.
\end{eqnarray*}

\begin{figure*}[t]
\footnotesize
\begin{tabular}{cc}
& $d_2$ \\
$d_1$ &
\begin{tabular}{|c||c|c|c|c|}
\hline
& $-\infty$ & $0$ & $k>0$ & $\infty$ \\ \hline \hline
\multirow{2}{*}{$-\infty$}
 &  & $X - (A,U) - B$ & & $X \perp B$ \\
 & $H(A) \geq I(X;A,B)$ & $H(A) \geq I(X;A,U) + I(A;U)$ & $H(A) \geq I(X;A,B) + I(A;B)$ & \\ \hline
\multirow{2}{*}{$0$}
 & $X \perp U, \quad H(A|X,U)=0$ & $X-A-B$ & $X \perp B$ & $X \perp B$ \\
 & $H(A) \geq I(X;A,B|U)$ & & & \\ \hline
\multirow{2}{*}{$k>0$}
 & $X \perp A$ & $X \perp (A,B)$ & $X \perp (A,B)$ & $X \perp (A,B)$ \\
 & $H(A) \geq I(X;A,B)$ & & & \\ \hline
\multirow{2}{*}{$\infty$}
 & $X \perp (A,B)$ & $X \perp (A,B)$ & $X \perp (A,B)$ & $X \perp (A,B)$ \\
 & & & & \\ \hline
\end{tabular}
\end{tabular}
\caption{The coordination set under various delay constraints.}
\label{fig.d1d2}
\end{figure*}

\begin{theorem}
\label{theorem causal}
The coordination set ${\cal P}$ for the cascade of controllers in Figure \ref{figure setup} with a strict causality constraint on Controller 2 is the set of conditional distributions $p(a,b|x)$ such that the joint distribution with the source, given by $p_0(x)p(a,b|x)$, satisfies
\begin{eqnarray*}
H(A|X,B) & \geq & I(X;B).
\end{eqnarray*}
\end{theorem}

\subsection{Achievability}

We use block-Markov coding.  Each block is of length $k$, and we denote the $i$th block $X^n$(i).  Consider a joint distribution $p_0(x)p(a,b|x)$ that satisfies $H(A|X,B) > I(X;B)$, and select $r$ such that $H(A|X,B) > r > I(X;B)$.  We generate a codebook ${\cal C}$ of $B^k$ sequences of size $2^{kr}$ according to the marginal distribution induced by  $p_0(x)p(a,b|x)$ to cover $X^k$.  We also randomly bin all the typical $A^k$ sequences in $2^{kr}$ bins.

At the beginning of  the $i$th block, Controller 1 finds an index $j_{i+1}$ in the codebook such that  $B^k(j_{i+1})$ is jointly typical with $X^n(i+1)$.  Controller 1 then finds an $A^k$ sequence in the $j_{i+1}$th bin that is jointly typical with $(X^n(i), B^k(j_i))$ and outputs that $A^k$ sequence in the $i$th block.  At the end of the $i$th block, Controller 2 observes the $A^k(i)$ sequence from Controller 1, thus decodes the bin index $j_{i+1}$.  In the $(i+1)$th block, Controller 2 simply outputs $B^k(j_{i+1})$ as its actions. This scheme works with high probability and yields an empirical distribution close to $p_0(x)p(a,b|x)$.

\subsection{Converse}

\vspace{-.3in}
{\allowdisplaybreaks
\begin{eqnarray*}
nH(X) &=& H(X^n)\label{eq.converse_d1=1}\\
&\stackrel{(a)}{=}& H(X^n,A^n)\nonumber\\
&=& \sum_{q=1}^n H(X_q, A_q| X^{q-1}, A^{q-1} )\nonumber\\
&\stackrel{(b)}{=}&\sum_{q=1}^n H(X_q, A_q| X^{q-1}, A^{q-1} , B_q)\nonumber \\
&\leq& \sum_{q=1}^n H(X_q, A_q| B_q)\nonumber\\
&=&  nH(X_Q, A_Q | B_Q,Q )\nonumber\\
&\leq& nH(X_Q, A_Q |B_Q) ,\nonumber
\end{eqnarray*}}
where (a) is because $A^n$ is a function of $X^n$ and (b) is due to the fact that $B_q$ is a function of $A^{q-1}$. The random variable $Q$ is uniformly distributed on the set $[n]$ and independent of $\{X^n,A^n,B^n\}$. Note that $H(X)\leq H(X,A|B)$ is equivalent to $I(X;B) \leq  H(A|X,B)$.

Based on Theorem~\ref{theorem causal} and the discussion in Section~\ref{subsection average score}, we can characterize the optimal average score of the game in the following corollary:

\begin{corollary}
For a game that pays out $\pi(x,a,b)$ ($x$ represents the source realization, $a$ represents the action of Alice, $b$ represents the action of Bob), and an i.i.d. source sequence with distribution $p_0(x)$, the optimal average score of the game (assuming Alice knows the entire source sequences, Bob sees past actions of Alice, and they produce actions simultaneously) is
\begin{equation*}
\max_{p(a,b|x) \; : \; H(A|X,B) \geq I(X;B)} \mathbf{E} \; \pi (X,A,B).
\end{equation*}
\end{corollary}

{\em Remark: } If we specialize the corollary to the case where $X\sim$Bernulli$(1/2)$ and carry out the optimization we will recover the optimal score in \cite{Gossner03}.  Furthermore, the score cannot be improved even if Bob is allowed to also see the past source realizations (that he has already attempted to guess).  The converse for Theorem \ref{theorem causal}, in particular inequality (b), still holds.

\section{Further Extensions}
In general, the encoding functions for both controllers can be subject to delay constraints, i.e,
\begin{eqnarray*}
A_i & = & f_i(X^{i-d_1}), \\
B_i & = & g_i(A^{i-d_2}),
\end{eqnarray*}
where $d_1$ and $d_2$ are the delays. The results under different $d_1$ $d_2$ combinations are listed in Figure~\ref{fig.d1d2}. Note that $-\infty$ means non-causal. Section~\ref{section setup} solved the case $d_1=-\infty$ and $d_2=-\infty$ and Section~\ref{section causal} solved the case $d_1 = -\infty$ and $d_2 = 1$.

\end{document}